\documentclass{elsart}
\usepackage{graphicx,amssymb,epsfig}
\journal{Physics Letters A}
\begin{document}
\begin{frontmatter}

\title{
Planar Dirac fermions in long-range-correlated \\ 
random vector potential}
\author[NC]{D.V. Khveshchenko\corauthref{cor}}, 
\corauth[cor]{Corresponding author.}
\ead{khvesh@physics.unc.edu}
\author[NC,SP]{A.G. Yashenkin}
\ead{ayash@thd.pnpi.spb.ru}

\address[NC]{Department of Physics and Astronomy, University 
of North Carolina,\\
 Chapel Hill, NC 27599, USA}
\address[SP]{Petersburg Nuclear Physics Institute, Gatchina, 
St. Petersburg 188300, Russia}

\begin{abstract}
We study the behavior of two-dimensional Dirac fermions in the presence 
of a static long-range-correlated random vector potential. By applying 
an exact path integral representation for the propagator of a spinor
particle we obtain asymptotics of the gauge invariant spectral function 
and the correlation function of the local density 
of states, both in the ballistic regime of sufficiently high energies.
We also discuss localization properties of the random Dirac wave 
functions in the complementary zero energy limit and the putative 
localization scenario.
\end{abstract}
\begin{keyword}
Dirac fermions, random magnetic field, vortex line liquid, Bohm-Aharonov phase  
\PACS 71.10.Pm \sep 72.10.-d
\end{keyword}
\end{frontmatter}

\section{Introduction}

In recent years, the standard theory of electron localization in two 
dimensions has been extended to the situations where, instead of potential 
disorder, the fermions are subject to a random vector potential (often 
referred to as random magnetic field or RMF) whose Gaussian spatial 
fluctuations are described by a pairwise correlator 
\begin{equation}
\label{eq:RMF}
\langle a_i({\vec q})a_j({\vec q^{\prime}}) \rangle =
w_{ij}(\vec q)\, 
\delta ({\vec q}+{\vec q^{\prime}}) \, ,
\end{equation}
where $w_{ij}(\vec q) =(\alpha/|{\vec q}|^{\eta})\left(\delta_{ij}-\frac{q_i q_j}{q^2}\right)$.

This problem was first encountered and then extensively studied in the context 
of the compressible Quantum Hall Effect at even denominator filling 
fractions. In order to describe the ostensibly Fermi liquid-like 
properties of these strongly correlated electronic states, a new kind of 
quasiparticles (``composite fermions'') which would effectively ``see'' 
an ordinary potential disorder as a long-range-correlated RMF described 
by Eq.(\ref{eq:RMF}) with $\eta=2$ was introduced in \cite{Zhang,HLR}.

Shortly thereafter, a (pseudo)relativistic version of the 
short-range-correlated, $\eta=0$, RMF problem had been proposed as the means 
of describing the Integer Quantum Hall plateau transitions 
\cite{Ludwig} where it was later identified with the continuum limit of 
some network models. The intrinsic scale invariance of the $\eta=0$ 
problem prompted the use of the powerful machinery of two-dimensional 
conformal field theory in most of the subsequent work on this topic.

Recently, a novel (this time, long-range correlated, $\eta=2$) variant 
of the relativistic RMF problem has emerged in the theory of 
localization of the Dirac-like nodal quasiparticles in the mixed state of
planar $d$-wave superconductors. 

In the presence of vortices of the $d$-wave superconducting order 
parameter, not only do the nodal quasiparticles' energies get 
semiclassically Doppler shifted due to the circulating supercurrent
\cite{Volovik},
but also their wave functions acquire intrinsically quantum mechanical
Bohm-Aharonov (BA) phases. 

In order to account for both effects on equal footing, a singular 
gauge transformation similar to that of Ref.\cite{HLR} was implemented in Ref.\cite{FT1}. 
In the case of a disordered vortex array, 
which corresponds to the experimentally well-documented vortex line liquid 
phase where the vortices are believed to be randomly pinned by columnar 
or other strong defects, the transformation of Ref.\cite{FT1} converts 
the nodal quasiparticles into auxiliary neutral Dirac fermions. The 
latter, in addition to a random scalar potential accounting for the 
Doppler shift, are subject to an effective RMF with zero mean 
described by Eq.(\ref{eq:RMF}) where $\eta=2$ and $\alpha$ is 
proportional to the areal density of vortices \cite{Yu,Franz,Vekhter,Ye}. 

Moreover, a formally similar RMF problem has emerged in the theory of the 
pseudogap phase of the cuprates described as a plasma of thermally excited 
vortex-antivortex pairs, the parameter $\alpha$ being proportional to 
temperature \cite{Millis,Dorsey}.  
Furthermore, one encounters yet another formal analog of this problem when
analyzing the effect of topological structural defects (dislocations)
on the Dirac-like electronic excitations with the momenta near 
conical ($K$-)points in the hexagonal Brillouin zone of a graphite layer  
\cite{Guinea}.

However, despite its occurrence in all of the above physical 
situations, the $\eta=2$ RMF problem for massless Dirac fermions
has, thus far, remained unsolved. In fact, this problem does not 
appear to be readily amenable to any of the methods applied to its previously 
studied $\eta=2$ non-relativistic \cite{Zhang,HLR} and $\eta=0$ Dirac
\cite{Ludwig} counterparts.

In the present paper, we fill in this gap by developing a novel 
non-perturbative path integral approach to the single-fermion action
\begin{equation}
\label{eq:Lagr}
{\rm S}=\int {\rm d}{\vec r}\int {\rm d}t{\overline \psi}(t,{\vec r})\left[ i{\hat 
\gamma}_0{\partial_t}-{\hat {\vec \gamma}}({\vec \nabla}-
{\vec a}({\vec r})) \right] \, \psi(t,{\vec r}) \, , 
\end{equation}
where the correlation properties of the static random vector 
potential ${\vec a}(\vec r)$ are governed by Eq.(\ref{eq:RMF}), and 
the ${\hat \gamma}_\mu$-matrices obey the algebra
${\{}{\hat \gamma}_\mu,{\hat \gamma}_\nu{\}}={\vec 1}
\delta_{\mu\nu}$ (hereafter, we put the speed of the Dirac fermions 
equal to unity).

\section{Path integral representation for gauge-invariant propagator}

In order to gain a preliminary insight into the problem we first 
attempt to apply the customary self-consistent Born approximation 
(SCBA) to the conventional (retarded) fermion propagator 
${\hat {\rm G}}^R (t,{\vec r})=\Theta (t)\langle\psi (t,{\vec r}) 
{\overline \psi} (0, {\vec 0})\rangle$. A closed SCBA equation can 
be obtained for the fermion self-energy which is defined via the Fourier 
transform
${\hat {\rm G}}^R (\epsilon, {\vec p})=[\epsilon {\hat \gamma}_{0} 
- {\vec p}\,{\hat {\vec \gamma}} + 
{\hat {\Sigma}}^R (\epsilon, {\vec p}) ]^{-1}$
\begin{equation}
\label{eq:SCBA}
{\hat {\Sigma}}^R (\epsilon, {\vec p})=
\int \frac{{\rm d} {\vec q}}{(2 \pi)^2} \, w_{ij}({\vec q})\, 
{\hat \gamma}_i {\hat {\rm G}}^R (\epsilon,{\vec p}+{\vec q}) 
{\hat\gamma}_j \, .
\end{equation}
Although one can readily solve Eq.(\ref{eq:SCBA}) for any $\eta<2$, 
including the previously studied case of $\eta=0$ \cite{Ludwig}, 
at $\eta=2$ the analysis is impeded by the fact that this equation
now exhibits an infrared divergence with the size $L$ of the system 
(${\hat \Sigma}\propto {\hat \gamma}_0 \sqrt{\ln L}$), thus 
rendering the SCBA inapplicable.

This intrinsic divergence remains for all $\epsilon$ and ${\vec p}$, 
even if one proceeds beyond the SCBA by replacing (\ref{eq:SCBA}) 
with the equation ${\hat \Sigma}=n\,{\hat {\rm T}}*{\hat {\rm G}}$, 
where $n$ stands for the density of Poissonian-distributed 
localized RMF sources (hereafter, motivated by the discussion of 
the mixed state of $d$-wave superconductors, we refer to them as 
``vortices''). Although the exact ${\rm T}$-matrix given by
the expression ${\hat {\rm T}}={\hat w}*({\hat I}-
{\hat {\rm G}}*{\hat w})^{-1}$ does account for all the 
events of multiple scattering by the same vortex, the above infrared 
divergence can not be made disappear, for it is  directly related to 
the non-gauge invariant nature of ${\hat {\rm G}}^R(t,{\vec r})$.

This problem does not arise, however, when calculating
such observables such as the gauge-invariant Green function
of the nodal quasiparticles introduced in 
in the context of the $d$-wave superconductors \cite{Wen,Ye2,FT2}
\begin{equation}
\label{eq:Gdef}
{\hat {\rm G}}^{R}_{{\rm inv}}(t,{\vec r})=\Theta(t)\langle
\psi(t,{\vec r})\,\exp\left(-i\int_{\Gamma}{\vec a}({\vec r}^\prime)
{\rm d} {\vec r}^\prime \right)\,\, {\overline \psi}(0,{\vec 0})
\rangle , \end{equation}
which is manifestly invariant under arbitrary time-independent 
gauge transformations for any choice of the contour $\Gamma$.

In Refs.\cite{Wen,Ye2,FT}, where the amplitude (4) was discussed in the case of 
a dynamic vector potential ${\vec a}(t, {\vec r})$ (the latter being
a part of the fictitious time-dependent gauge field representing 
either spin \cite{Wen} or pairing \cite{Ye2,FT} fluctuations), the contour $\Gamma$ was chosen
as a straight line between the end points $(0, {\vec 0})$ and $(t, {\vec r})$.

It must be noted, however, that in the Lorentz-invariant situation
considered in Refs.\cite{Wen,Ye2,FT} the amplitude (4) with the straight-line
contour $\Gamma$ appears to exhibit an unphysical power-law
behavior characterized by a negative (instead of a positive, as 
one would have expected on the general physical grounds and as Refs.\cite{Wen,Ye2,FT}
erroneously claimed to be the case) anomalous exponent \cite{DVK}.

Although, thus far, no physically sound alternative to the heuristic
form of Eq.(4) conjectured in Refs.\cite{Wen,Ye2,FT} has been found, it turns out that
in the non-Lorentz-invariant case of a static vector potential 
${\vec a}({\vec r})$ the situation is rather different.

Namely, in the case of interest, the most important
features, such as the mean free path (see Eqs.(11,12) below) which controls the exponential 
(rather than a power-law, as in the case of a dynamic RMF \cite{Wen,Ye2,FT,DVK}) 
decay of Eq.(4) or the structure of the "near-shell" singularity
of the fermion spectral function (see Eqs.(13,14) below), remain robust against
deformations of the contour $\Gamma$ (apparently, in the static case
the only potentially important is the spatial projection
of $\Gamma$, whereas its pitch in the temporal direction is not).

Moreover, while being representative of the properties of a generic gauge-invariant
one-fermion amplitude, the above choice of the contour $\Gamma$ 
facilitates the calculation of Eq.(4) along the lines of the earlier analyses 
of the non-relativistic version of the $\eta=2$ RMF problem \cite{AI,KM,Mirlin1,Mirlin2}.

Of course, the spinor nature of the amplitude (4) makes it impossible
to apply the results of Refs.\cite{AI,KM,Mirlin1,Mirlin2} directly.
Therefore, we first utilize the (relatively unknown) path integral representation 
devised in \cite{Stefanis2} and, for a fixed RMF configuration 
${\{}{\vec a}({\vec r}){\}}$, cast the time Fourier transform of  
Eq.(4) in the form of a path integral over the spatial
coordinate ${\vec r}(\tau)$ and an auxiliary vector-like variable 
${\vec k}(\tau)$ parametrized by the proper time $\tau$ (not to 
be confused with the real time $t$!)
\begin{equation}
\label{eq:Gcont}
{\hat {\rm G}}^{R}_{{\rm inv}} (\epsilon,{\vec r} | {\vec a})=
i \int^{\infty}_{0} {\rm d}\tau
\int_{{\vec r}(0)={\vec 0}}^{{\vec r}(\tau)={\vec r}}
{\rm D} {\vec r} \, {\rm D} {\vec k} \,
\exp \left( i{\hat {\rm S}}_0(\tau)+i\int_0^{\tau}{\vec a}({\vec r})
\frac{{\rm d}{\vec r}} {{\rm d} \tau^\prime} {\rm d} \tau^{\prime}
-i\int_{\Gamma}{\vec a}({\vec r}^\prime){\rm d} {\vec r}^\prime
\right)
\end{equation}
where the first term in the exponent represents the (matrix-valued) free fermion action
\begin{equation}
\label{eq:Sbare}
{\hat {\rm S}}_0 (\tau)=\int_{0}^{\tau} {\rm d} \tau^{\prime} 
\left[ \epsilon {\hat \gamma}_{0} + {\vec k} \left( 
\frac{{\rm d} {\vec r}}{{\rm d} \tau^{\prime}} -
{\hat {\vec \gamma}} \right)\right] \, , 
\end{equation}
while the other two correspond to the RMF coupling term in 
Eq.(\ref{eq:Lagr}) and the line integral in the exponential factor inserted into  
Eq.(\ref{eq:Gdef}), respectively. In Eq.(\ref{eq:Gcont}), the usual 
$\tau$-ordering of the ${\hat \gamma}$-matrices must be performed 
according to the order of their appearance in the series expansion of the 
exponent. In contrast to various approximate (e.g., 
Bloch-Nordsieck) representations, the integration over 
${\vec k}(\tau)$ allows one to exactly account for the spinor 
structure of the fermion propagator.

Averaging (\ref{eq:Gcont}) over the Gaussian RMF introduces the 
exponential attenuation factor
\begin{equation}
\label{eq:Wdef}
W[{\vec r}(\tau)]=\exp
\left[-\frac{1}{2}\int^\tau_0 {\rm d} \tau_1 \int^\tau_0 {\rm d} \tau_2
\int {{\rm d}{\vec q}\over (2\pi)^2} \,
e^{i{\vec q}({\vec r}(\tau_1)-{\vec r}(\tau_2))}\,
w_{ij}({\vec q})\, \frac{{\rm d} r_i}{{\rm d}\tau_1}
\frac{{\rm d} r_j}{{\rm d}\tau_2} \right] \, ,
\end{equation}
whose argument, in the case of $\eta=2$, turns out to be
proportional to the so-called Amperian area of a  
closed contour which is composed of a given trajectory 
${\vec r}(\tau)$ and the straight-line segment traversed 
backwards (from $\vec r$ to $\vec 0$) \cite{AI,KM}.
 
\section{Gauge-invariant propagator in the ballistic regime}

While being strictly positive, the Amperian area coincides with the 
absolute value of the regular (algebraic) one only for 
non-self-intersecting closed contours, which, strictly speaking, 
precludes Eq.(\ref{eq:Wdef}) from being interpreted as an 
additional term in the effective Dirac action.
Nevertheless, this becomes possible for the trajectories whose 
projections onto the straight path $\Gamma$ are single-valued, 
whence (\ref{eq:Wdef}) reduces to
\begin{equation}
\label{eq:Wperp}
W[{\vec r}(\tau)]\approx\exp\left(-\frac{\alpha}{2} 
\int^r_0 |x_\perp|\, {\rm d} x_\parallel \right) \, ,
\end{equation}
where the parallel and transverse components of the vector 
${\vec r}(\tau)=(x_{\parallel},x_{\perp})$ are defined with respect 
to the direction of the fermion's propagation.

The relativistic kinematics of the Dirac fermions makes the 
present situation markedly different from the previously studied 
non-relativistic version of the $\eta=2$ RMF problem 
\cite{AI,KM,Mirlin1,Mirlin2}.
As opposed to the interpretation of the non-relativistic
analog of $\delta {\rm S}=i\ln W[{\vec r}(\tau)]$ as an effective linear 
confining potential \cite{AI,KM,Mirlin1,Mirlin2}, 
Eq.(\ref{eq:Wperp}) can be best 
thought of as an (imaginary) position-dependent mass whose linear 
increase with $|x_\perp |$ restrains the Dirac fermion's motion
in the direction perpendicular to its ``classical trajectory'' 
${\vec r}_0(\tau)=\tau ({\vec r}/r)$, the latter representing the saddle point 
of the path integral (5).

Comparing the saddle-point action 
${\rm S}_0\sim\epsilon\tau$ to the characteristic RMF correction
$\delta {\rm S}\sim(\alpha\epsilon/p)^{1/2}\tau$, we 
find that the condition ${\rm S}_0\gg\delta {\rm S}$, under which the 
path integral (\ref{eq:Gcont}) is dominated by the trajectories 
close to ${\vec r}_0(\tau)$, is satisfied in the 
``ballistic'' regime of high energies and large momenta 
($\epsilon p\gg{\alpha}$).

In this regime, one can integrate over the momentum 
${\vec p}(\tau)$, thereby effectively squaring the Dirac operator 
(for details, see \cite{DVK,Stefanis2}), as well as the 
longitudinal spatial $x_\parallel(\tau)$ component of the 
coordinate vector ${\vec r}(\tau)$, thus reducing 
(\ref{eq:Gcont}) to the one-dimensional path integral over 
$x_\perp(\tau)$.

As in the non-relativistic $\eta=2$ RMF 
problem \cite{AI,KM,Mirlin1,Mirlin2}, the computation of the remaining 
path integral amounts to finding the resolvent of the second order 
ordinary differential equation
\begin{equation}
\label{eq:diffur}
\left[\, ( p_{\parallel}^{2}- \epsilon^2 ) - \partial^2_{\perp} + 
\left(\alpha \epsilon / p_{\parallel}\right)^2 x^{2}_{\perp}
+ i{\alpha} \, {\rm sign}\, x_\perp \right] \, g(\epsilon, 
p_\parallel | x_\perp,
x_{\perp}^\prime)=\delta(x_\perp-{x_{\perp}^\prime}) \, ,
\end{equation}
whose exact form can be readily obtained in terms of the 
Wronskian of a pair of linearly independent solutions of the 
corresponding homogeneous equation 
\begin{equation}
\label{eq:solution}
g(\epsilon, p_\parallel | 0,0)=
{\sqrt \frac{i p_\parallel}{\alpha |\epsilon|}}
\left[
\sum_{\lambda=\pm} \frac{\Gamma(z_\lambda+1/2)}{\Gamma(z_\lambda)}
\right]^{-1}\, ,
\end{equation}
where 
$z_\lambda=i p_\parallel (\epsilon-\lambda p_\parallel)
(\epsilon +\lambda p_\parallel+i\alpha/p_\parallel)
/4\epsilon\alpha$ and $\Gamma(z)$ is the standard 
$\Gamma$-function.

The asymptotical large-distance behavior of the gauge-invariant 
propagator is controlled by the closest to the 
real $p_{\parallel}$-axis zero of the denominator in 
Eq.(\ref{eq:solution}). Applying to 
$g(\epsilon, p_{\parallel} | 0,0)$ the one-dimensional Fourier 
transformation with respect to $p_{\parallel}$ we arrive at the 
expression
\begin{equation}
\label{eq:largeR}
{\hat {\rm G}}^{R}_{{\rm inv}}(\epsilon,{\vec r})\sim
\left(\epsilon{\hat \gamma}_0-\frac{{\vec r}}{r^2}{\hat {\vec \gamma}} 
\right){e^{-r/l}\over {\sqrt {|\epsilon| l}}}
\sin(|\epsilon|r+\phi_1) ,
\end{equation}
where $\phi_1\sim 1$, which decays exponentially at $r \to \infty$
with the characteristic length scale
\begin{equation}
\label{eq:MFP}
l\approx \frac{8|\epsilon|}{\pi\alpha}  
\end{equation}
The latter can be naturally identified with the RMF mean free path
which, according to Eq.(12), decreases as the fermion energy 
gets lower, contrary to the situation for $\eta=0$ \cite{Ludwig} 
(but qualitatively similar to that found in the non-relativistic 
variant of the $\eta=2$ problem where 
$l\sim |\epsilon|^{1/6}\alpha^{-2/3}$ \cite{AI,KM,Mirlin2}).

The ``near-maximum'' ($|{\epsilon}^2 - p^2| < \alpha$) behavior of 
the electron spectral function is determined by the two-dimensional 
Fourier transform of ${\rm Im}\,{\hat {\rm G}}^{R}_{{\rm inv}} 
(\epsilon, {\vec r})$
which can be obtained by convoluting Eq.(\ref{eq:solution}) with the kernel 
$(p^2-q^2)^{-3/2}$, thus resulting in the expression
\begin{equation}
\label{eq:close}
{\rm Im}\,{\hat {\rm G}}^{R}_{{\rm inv}} (\epsilon, {\vec p})\approx 
\sqrt{\frac{\epsilon}{l}}
\frac{\epsilon {\hat \gamma_0}-{\vec p}\,{\hat {\vec \gamma}}}
{[(\epsilon^2-p^2)^2+(\epsilon/l)^2]^{3/4}}\, ,
\end{equation}
which explicitly demonstrates that the bare pole gets replaced
by branch cuts of the function $z_\lambda^{3/2}$
(hence the emergence of the exponent 3/4 in (\ref{eq:close})).

At larger deviations from the maximum 
($\alpha\ll|\epsilon^2-p^2| < \epsilon^2$) the spectral 
function is determined by the large-$z_\lambda$ asymptotics of
the $\Gamma$-functions in Eq.(\ref{eq:solution}). In this
limit, the spectral function resembles the Lorentzian
\begin{equation}
\label{eq:away}
{\rm Im}\,{\hat {\rm G}}^R_{{\rm inv}}(\epsilon,{\vec p}) \propto 
\alpha \frac{\epsilon {\hat \gamma_0}-{\vec p}\,{\hat {\vec \gamma}}}
{(\epsilon^2-p^2)^2 + \alpha^2} \, ,
\end{equation}
although, under a closer inspection, the shape of 
${\rm Im}\,{\hat {\rm G}}^{R}_{{\rm inv}}(\epsilon,{\vec p})$ turns out to be 
asymmetrical as a result of the Lorentz invariance's being 
broken by the static disorder. Moreover, the product of its 
maximum height and width reveals a marked departure from the 
Lorentzian value ($\approx\pi/4$ instead of $1$).

\section{Averaged products of Green functions}

Our eikonal-type approach also enables one to study 
other gauge-invariant physical observables, such as a correlation 
function of the local density of states (DOS) which is related to 
the average $\langle{\hat {\rm G}}^{R}
{\hat {\rm G}}^{R}\rangle$. 
Owing to its gauge invariant nature, the latter is free of 
infrared divergencies and can be cast in the form of a 
two-particle path integral
\begin{eqnarray}
\label{eq:GRGR}
\langle{\hat {\rm G}}^{R}(\epsilon, {\vec r})
{\hat {\rm G}}^{R}(\epsilon, -{\vec r})\rangle
=\int^\infty_0 {\rm d} \tau_1\int^\infty_0 {\rm d} \tau_2
\int_{0}^{{\vec r}}{\rm D} {\vec r}_1 {\rm D} {\vec k}_1 
\int_{0}^{{\vec r}} {\rm D}{\vec r}_2 {\rm D} {\vec k}_2
\cr
e^{i{\hat {\rm S}}_0(\tau_1)}\, e^{i{\hat {\rm S}}_0(\tau_2)}\,
\prod_{i,j =1,2}W({\vec r}_{i}(\tau_1)-{\vec r}_{j}(\tau_2))\, ,
\end{eqnarray}
where the product of the $W$-factors yields the exponent of 
the Amperian area of the closed contour formed by a pair of 
trajectories ${\vec r}_1(\tau)$ and ${\vec r}_2(\tau)$.

We mention, in passing, that the average (15) equals that composed of 
a pair of gauge-invariant amplitudes ($\langle{\hat {\rm G}}^{R}
{\hat {\rm G}}^{R}\rangle=\langle{\hat {\rm G}}^{R}_{{\rm inv}}
{\hat {\rm G}}^{R}_{{\rm inv}}\rangle$), since, regardless of the choice of 
$\Gamma$ in Eq.(4), the phase factors
from the two functions ${\hat {\rm G}}^{R}_{{\rm inv}}$ exactly 
cancel against each other.

In the ballistic regime, the integral (15) receives its main 
contribution from the pairs of trajectories with single-valued 
projections to the direction of the vector $\vec r$. Upon separating the 
coordinate variables onto the ``center of mass'' and the relative 
motion (${\vec r}^{\pm}={\vec r}_1 \pm {\vec r}_2$,
${\vec k}^{\pm}={\vec k}_1\pm {\vec k}_2$) parametrized by
$\tau^{\pm}=\tau_1\pm\tau_2$ and integrating over all the 
variables but $x^{-}_\perp(\tau^{+})$, one finds that 
the average (\ref{eq:GRGR}) can be again related to the solution of
Eq.(\ref{eq:diffur}), albeit with some extra powers of two stemming 
from the Jacobian of the above transformation. As a result, 
the variance of the local DOS $(\delta\nu(\epsilon,{\vec r})=
\nu(\epsilon,{\vec r})-\langle\nu(\epsilon)\rangle)$ 
decays with distance as 
\begin{equation}
\label{eq:CorNu}
\langle\delta\nu(\epsilon,{\vec r})\delta\nu(\epsilon, {\vec 0})
\rangle\propto \frac{|\epsilon | e^{-2r/l}}{\sqrt{rl}}
\sin(2|\epsilon|r+\phi_2)\, ,
\end{equation}
where $\phi_2\sim 1$. 

Notably, the average of the product of two Green functions (16)
does not amount to the product of the two averages (each of which
is given by Eq.(11)), thereby indicating the presence of non-trivial vertex corrections. 

Nevertheless, the exponential decay of Eq.(16)
is controled by a length scale equal to the 
half of the mean free path (12), which corroborates our
conclusion that (alongside a number of other robust features
such as the "near-shell" asymptotic behavior (13))
Eq.(12) is, in fact, independent of the choice of the contour $\Gamma$ in Eq.(4).

In a similar manner, one finds the average
$\langle {\hat {\rm G}}^{A}{\hat
{\rm G}}^{R}\rangle=\langle {\hat {\rm G}}^{A}_{{\rm inv}}{\hat
{\rm G}}^{R}_{{\rm inv}}\rangle\propto(\epsilon^2/lr)^{1/2}e^{-2r/l}$
which can be used for computing such transport 
coefficients as optical conductivity $\sigma(\omega, T)$. However, such a  
result would be limited to the regime of high frequencies or 
temperatures $\max(\omega, T)\gg \alpha^{1/2}$ where other mechanisms
of scattering, which may exist in the known realizations of the RMF problem 
(see Refs.\cite{Yu,Franz,Vekhter,Ye,Millis,Dorsey,Guinea}), might become 
important.

\section{Localization of zero-energy modes}

At low energies and small momenta $\epsilon p < \alpha $ our 
approach ceases to be valid, and, for instance, the low-energy 
behavior of the DOS can not be readily inferred from the above 
discussion. In order to shed some light on the effect of RMF 
scattering on the low-energy part of the Dirac spectrum we look 
into the properties of the zero-energy states. Quite remarkably, the 
latter exist for an arbitrary configuration of the non-uniform 
magnetic field, thanks to the exact cancellation between the 
orbital and the Zeeman terms in the total energy of a spinor particle 
with the gyromagnetic ratio equal two.

Parametrizing an arbitrary (up to a gauge transformation) RMF 
configuration in terms of a scalar function 
$a_i ({\vec r})=\epsilon_{ij}\partial_j\Phi({\vec r})$ 
one can write a pair of independent zero-energy wave functions (there 
are no other states with $\epsilon=0$ as long as the total RMF 
flux vanishes) as follows
\begin{equation}
\label{eq:Psi}
\Psi_{\pm}({\vec r})\propto ({\vec 1}\pm {\hat \gamma}_0)
\pmatrix{e^{\Phi({\vec r})}
\cr
e^{-\Phi({\vec r})}}\, .
\end{equation}
In a finite-size system, the degree of the wave functions' 
localization (or a lack thereof) can be inferred from the set of 
inverse participation ratios
\begin{equation}
\label{eq:IPRdef}
{\rm P}_n=
{\langle
\frac{\int |\Psi({\vec r})|^{2n}{\rm d} {\vec r}}{L^2(\int 
|\Psi({\vec r}^\prime)|^2
{\rm d} {\vec r}^\prime )^n} \rangle} \, .
\end{equation}
Performing the Gaussian average over the disorder field 
$\Phi({\vec r})$ with the weight $P[\Phi({\vec r})]\propto
\exp(-\int {\rm d} 
{\vec r}({\vec \nabla}^2\Phi)^2/2\alpha)$ we obtain
\begin{equation}
\label{eq:IPR}
{\rm P}_{n}\propto \frac{\alpha^{n-1}}{L^{2}} \, ,
\end{equation}
which is suggestive of strong localization of the zero-energy states, 
the ``localization length'' being of order $\xi\sim\alpha^{-1/2}$. This
conclusion is in stark contrast with the $\eta=0$ case \cite{Ludwig} 
where the prelocalized zero-energy wave functions exhibit a multifractal 
spectrum of anomalous dimensions ${\rm P}_n\propto L^{-an+bn^2}$.
Conceivably, as the energy decreases past $\alpha^{1/2}$, the elastic 
mean free path (given by Eq.(\ref{eq:MFP}) at high energies) 
saturates at $l\sim\xi$, followed by the onset of localization at still 
lower energies.

In the $\eta=0$ case, it was recently argued \cite{ASZ,Altland} that 
the corresponding localization scenario belongs to the so-called 
``${\rm C}$''
universality class which encompasses the $d$-wave superconductors with 
a strong inter-node scattering in the absence of time reversal symmetry 
\cite{Zirnbauer1,Zirnbauer2}. It was also conjectured in 
Refs.\cite{ASZ,Altland} that 
the low-energy properties of the system are described by the non-linear
$\sigma$-model (NL$\sigma$M) on the coset space 
${\rm O}{\rm Sp}(2|2)/{\rm G}{\rm L}(1|1)$.

In order to extend this conjecture to the Dirac $\eta=2$ RMF problem one would 
have to derive an effective NL$\sigma$M of the appropriate symmetry by 
carefully separating between the massless and massive modes of the 
(this time, non-local) Hubbard-Stratonovich disorder fields, by analogy 
with the procedure carried out in the nonrelativistic $\eta=2$ RMF problem \cite{Efetov}.

There is a good reason to believe that the resulting localization 
scenario falls into the already existing classification of the $d$-wave
universality classes chartered in Refs.\cite{Zirnbauer1,Zirnbauer2}. 
However, given 
the notorious sensitivity of the $d$-wave localization patterns not 
only to the type (potential vs magnetic vs extended defects, etc.) but 
even to the strength (Gaussian vs unitary) of disorder, such important 
details as the asymptotical form of the low-energy DOS 
($\langle\nu(\epsilon)\rangle\propto\epsilon^2$ 
for a generic class ``${\rm C}$'' system \cite{Fisher})
or a possible crossover from ``${\rm C}$'' to another, e.g.,  
the so-called ``${\rm A}$'', universality class at intermediate energies due to the
predominantly small-angle nature of the BA scattering, remain to be worked out.

\section{Summary}

To conclude, in the present paper we investigated the properties of the 
two-dimensional Dirac 
fermions subject to a long-range-correlated random vector potential. In 
the ballistic regime of large quasiparticle energies, we obtained the 
asymptotics of a representative gauge-invariant fermion Green function and the DOS 
correlation function. In the complementary low-energy regime, we found a 
signature of strong localization of the zero-energy states by computing 
their inverse participation ratios.

Thus far,
it has been rather difficult to find a conclusive experimental evidence 
of the quasiparticle localization in the high-$T_c$ cuprates. However, 
theoretical predictions based on the results of this paper which 
pertain to the more experimentally accessible ballistic 
regime may be possible to test with a number of standard probes
(thermal transport, ARPES, tunneling, specific heat, NMR) performed in the 
vortex line liquid phase of the superconducting cuprates \cite{KY}. 

To this end, it is worth mentioning that the results of a recent numerical 
analyis of the low-field thermal quasiparticle conductivity in
the vortex line liquid state of the $d$-wave 
superconductors \cite{Durst} appear to be in good agreement 
with the analytical result based on the analog of Eq.(12) \cite{KY}. 

Moreover, the results of this paper can be used to study the 
effects of both thermal \cite{Millis,Dorsey} and quantum \cite{Wen,Ye2,FT} 
phase fluctuations on the spectrum of the nodal quasiparticles in the 
pseudogap phase of the cuprates as well as that of randomly distributed 
dislocations on the Dirac-like electronic excitations in graphite 
\cite{Guinea}. 

In particular, when applied to the theories of the pseudogap 
phase \cite{Wen,Ye2,FT} where the nodal fermions
are scattered by the quasi-static (thermal) gauge fluctuations which represent
either the effect of a disordered flux-phase \cite{Wen} or
that of thermally excited vortices \cite{Ye2,FT}, 
the results of this paper imply that the gauge-invariant propagator 
should exhibit an exponential, rather 
than a power-law, decay, contrary to the conclusions drawn in Refs.\cite{Wen,Ye2,FT}.

\noindent
{\bf Acknowledgments}

\noindent
This research was supported by the NSF under Grant No. DMR-0071362.

\end{document}